\documentclass[sigconf]{acmart}
\AtBeginDocument{%
 \providecommand\BibTeX{{%
  \normalfont B\kern-0.5em{\scshape i\kern-0.25em b}\kern-0.8em\TeX}}}

\usepackage{subfig}

\usepackage[utf8]{inputenc}
\usepackage[english]{babel}
\usepackage{algorithmic}

\usepackage[ruled,vlined,]{algorithm2e}

\copyrightyear{2021}
\acmYear{2021}
\setcopyright{acmcopyright}\acmConference[SIGGRAPH '21 Emerging Technologies ]{Special Interest Group on Computer Graphics and Interactive Techniques Conference Emerging Technologies }{August 9--13, 2021}{Virtual Event, USA}
\acmBooktitle{Special Interest Group on Computer Graphics and Interactive Techniques Conference Emerging Technologies (SIGGRAPH '21 Emerging Technologies ), August 9--13, 2021, Virtual Event, USA}
\acmPrice{15.00}
\acmDOI{10.1145/3450550.3465346}
\acmISBN{978-1-4503-8364-6/21/08}

\begin{document}

\title{SwarmPlay: A Swarm of Nano-Quadcopters Playing Tic-tac-toe Board Game against a Human}
\author{Ekaterina Karmanova}
\affiliation{%
 \institution{Skolkovo Institute of Science and Technology}
 \streetaddress{Nobel street 3}
 \city{Moscow}
 \country{Russia}}
\email{ekaterina.karmanova@skoltech.ru}

\author{Valerii Serpiva}
\affiliation{%
 \institution{Skolkovo Institute of Science and Technology}
 \streetaddress{Nobel street 3}
 \city{Moscow}
 \country{Russia}}
\email{valerii.serpiva@skoltech.ru}

\author{Stepan Perminov}
\affiliation{%
 \institution{Skolkovo Institute of Science and Technology}
 \streetaddress{Nobel street 3}
 \city{Moscow}
 \country{Russia}}
\email{stepan.perminov@skoltech.ru}

\author{Roman Ibrahimov}
\affiliation{%
 \institution{Skolkovo Institute of Science and Technology}
 \streetaddress{Nobel street 3}
 \city{Moscow}
 \country{Russia}}
\email{roman.ibrahimov@skoltech.ru}

\author{Aleksey Fedoseev}
\affiliation{%
 \institution{Skolkovo Institute of Science and Technology}
 \streetaddress{Nobel street 3}
 \city{Moscow}
 \country{Russia}}
\email{aleksey.fedoseev@skoltech.ru}

\author{Dzmitry Tsetserukou}
\affiliation{%
 \institution{Skolkovo Institute of Science and Technology}
 \streetaddress{Nobel street 3}
 \city{Moscow}
 \country{Russia}}
\email{d.tsetserukou@skoltech.ru}

\renewcommand{\shortauthors}{Karmanova, Serpiva, Perminov, Ibrahimov, Fedoseev and Tsetserukou }

\begin{abstract}
We present a new paradigm of games, i.e. SwarmPlay, where each playing component is presented by an individual drone that has its own mobility and swarm intelligence to win against a human player. The motivation behind the research is to make the games with machines tangible and interactive. Although some research on the robotic players for board games already exists, e.g., chess, the SwarmPlay technology has the potential to offer much more engagement and interaction with a human as it proposes a multi-agent swarm instead of a single interactive robot. The proposed system consists of a robotic swarm, a workstation, a computer vision (CV), and Game Theory-based algorithms. A novel game algorithm was developed to provide a natural game experience to the user.
The preliminary user study revealed that participants were highly engaged in the game with drones (69\% put a maximum score on the Likert scale) and found it less artificial compared to the regular computer-based systems (77\% put maximum score). The affection of the user's game perception from its outcome was analyzed and put under discussion. User study revealed that SwarmPlay has the potential to be implemented in a wider range of games, significantly improving human-drone interactivity. 
\end{abstract}



\begin{teaserfigure}
\centering
 \includegraphics[width=1\linewidth]{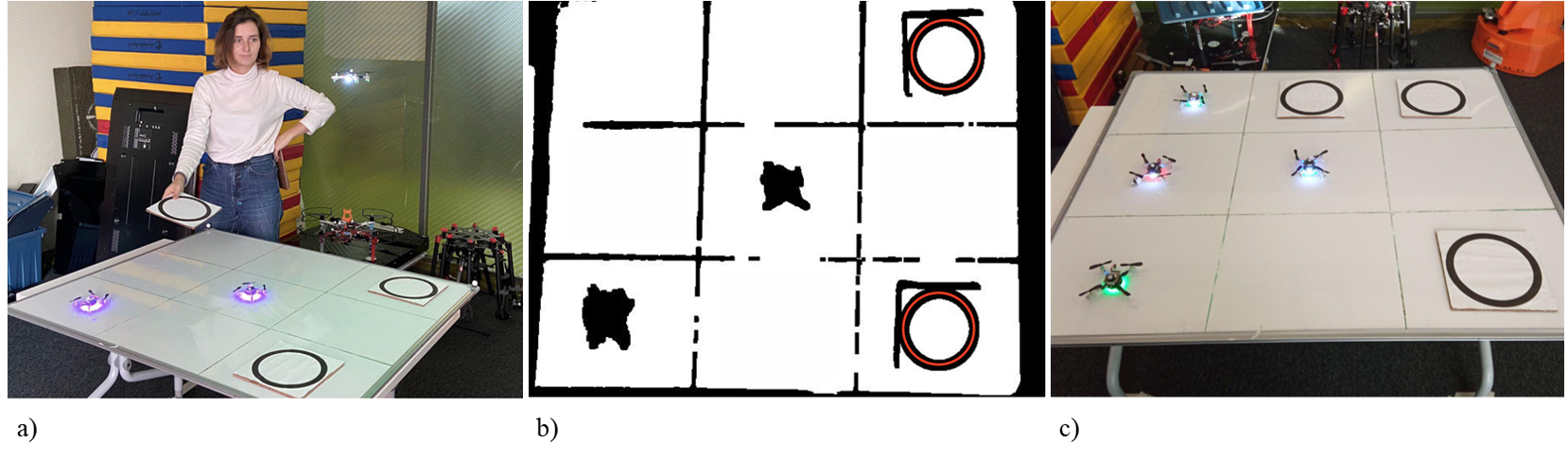}
 \caption{Tic-tac-toe game with the swarm of drones: a) A human plays Tic-tac-toe board game against a SwarmPlay. b) Game board image transformed by the CV system. c) A closed view on the board when drones won the match.}
 \label{fig:teaser}
\end{teaserfigure}

\maketitle

\section{Introduction}
One of the potential domains for human-robot interaction research is physical board games that have an adjustable structure level. Perceiving the game components and the board, understanding human movements, reasoning about the state, and manipulating the game components to win against human players are integral steps in robot-centric board games \cite {Li_Differential, matuszek2011gambit}. 

For the human player, on the other side, the interaction with a robot provides a fresh perspective on the well-known competitive games, e.g., robotic rock-paper-scissors with 
humanoid robot RASA presented in \cite{Ahmadi_Rock_Paper}.
Nowadays much work is aimed to improve the AI in such games, either by more sufficient game strategy systems, e.g., gaming decision system developed for Go in \cite{Silver2016MasteringTG} and curling robot with adaptive deep reinforcement learning framework proposed in \cite{Woneabb9764}, or by improving the estimation of the human behavior, e.g., tennis player's movement prediction proposed in \cite{Wu_FuturePong}. However, the system architecture in robot-centric applications has been relatively little investigated and is narrowed to the single robotic manipulators and mobile robots \cite{nugroho2014design,Kyohei_PP_Robot,Becker_Chess}. The research on multi-robot games though is mostly focused on coordination between robotic agents, such as soccer game strategies suggested in \cite{Reis_multiSoccer} and \cite{Liu_Soccer} that exclude human from the gaming stage.

 To upgrade the level of engagement and interactivity of traditional games, we suggest a novel game paradigm where each game piece has its own mobility, and behaves jointly with other agents to win against the opponent. The proposed SwarmPlay technology provides CV-driven Human-Swarm interaction (HSI) in board games. To our knowledge, our prototype is the first approach towards using a multi-UAV system in physical games that involves human presence. This research focuses on the system architecture and its validation by user study, followed by a discussion about future work and potential SwarmPlay game applications. 

\section{SYSTEM OVERVIEW}
\subsection{System Architecture} 

\begin{figure} [h]
\centering
\includegraphics[width=0.92\linewidth]{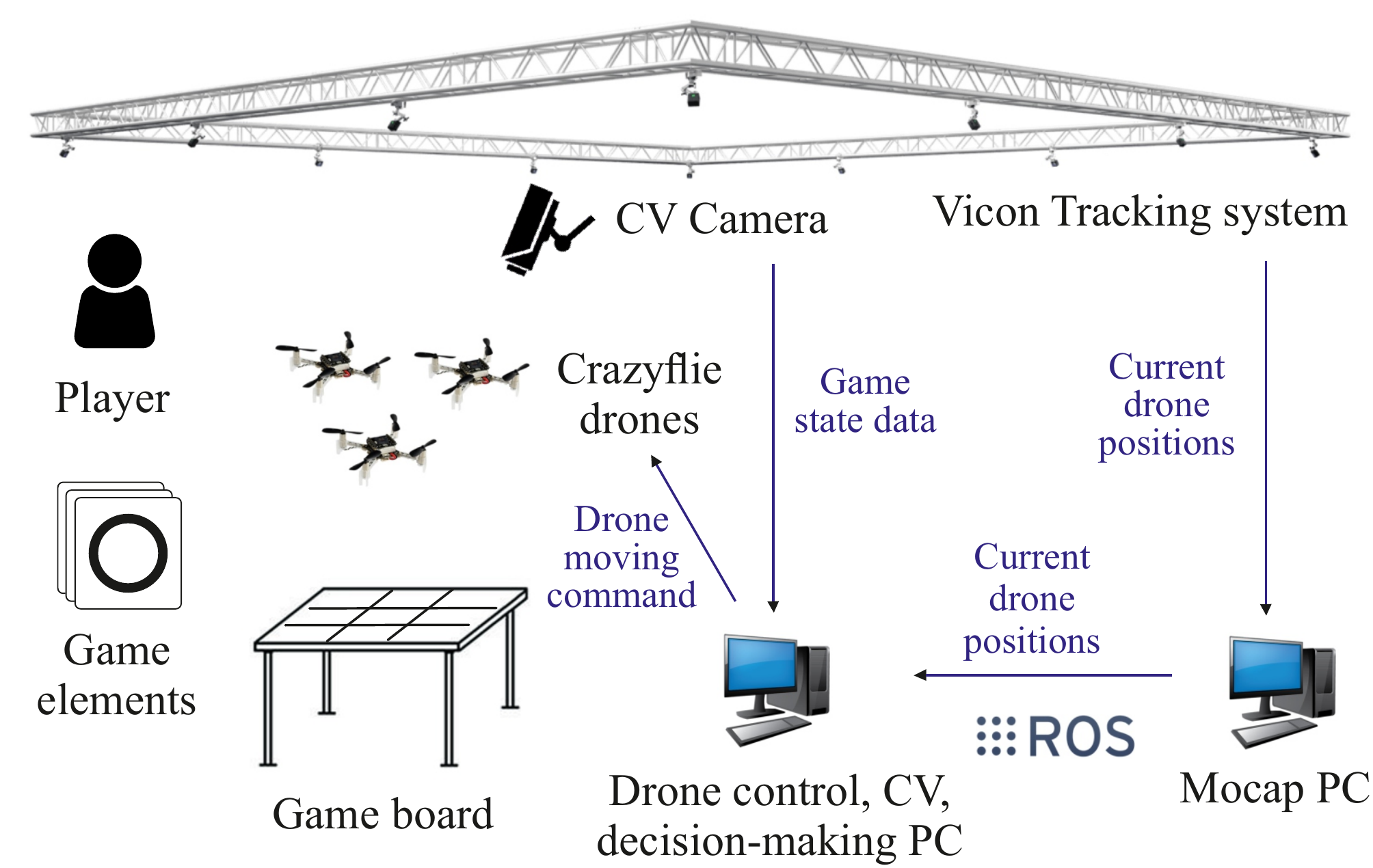}
 \caption{SwarmPlay system layout: hardware architecture.}
 \label{fig:Overwiev}
 \vspace{-0.5 em}
\end{figure}

The developed SwarmPlay system consists of Vicon Tracking system with 12 IR cameras for drone localization, a CV camera for game state evaluation, a drone landing table with a game board, PC with Mocap framework and PC with a drone-control framework, a CV system, and a decision-making system (Fig. \ref{fig:Overwiev}). A white board was divided into 9 cells according to Tic-tac-toe rules. According to the specified algorithm, drones were landing on a board's cells representing Crosses (Xs). Whereas a human plays Noughts (Ox), placing cards with printed circles on the white board.

To obtain pictures of the game board providing awareness of a current status of the game, we used a camera Logitech HD Pro Webcam C920 of @30FPS mounted on the ceiling of the room. The game board is placed right under the camera. The pictures are sent to the CV system to determine the human's turn. After that, data on the human's turn as a cell number is sent to the decision-making system to define a cell where the drones should make their next turn. CV and decision-making processing is performed on Intel® Core™ i7-9750HF CPU @ 2.60GHz × 12. The most recent cell data is sent to a drone-control framework. The framework obtains both the target cell, where a drone should be sent, and data from the motion capture system about current drone positions. To get the high-quality tracking of the drones, we used Vicon motion capture system with 12 cameras (Vantage V5) covering a ${5 m^3}$ space. Drones are sent to the target cells with PID control parameters, i.e., the target position, speed, and acceleration.

\subsection{Computer Vision}
To detect to which cell the user puts circles, we developed a corresponding Computer Vision system. As its input, we used a picture taken by an RGB camera mounted on the ceiling of the room where the gaming board is located. At each step, the system takes a picture throughout the camera and converts it to grayscale. Then, simple thresholding and erosion with a kernel 5x5 are applied.
After that, the picture is cropped and divided into 9 small images, one per a game cell. For each small image, a contour search is performed. When users make their turn, they put a circle on a cell, which is then detected as a contour by the CV system and filled with black pixels.

At the end of each step, the CV system computes the density of black pixels per a game cell. In this case, big coloured circles show a great density value. Thus, using some threshold, it makes possible to separate holes, drones, and empty areas from each other.

After detecting a new circle on the playing field, the CV system sends a corresponding game cell number, as the latest human turn, to a decision-making system to solve how exactly drones should behave in the situation.

\section{Game strategy}

\subsection{Implementation}

\begin{figure} [!h]
\centering
\includegraphics[width=0.8\linewidth]{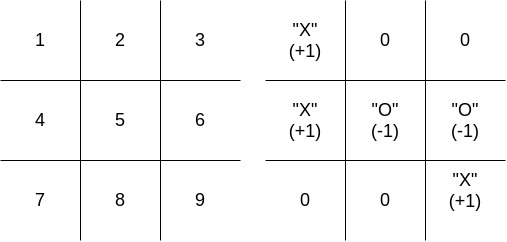}
 \caption{Game board representation in the SwarmPlay system.}
 \label{fig:board}
\end{figure}

Tic-tac-toe game is played on a three-by-three grid. Each player takes a turn to place a symbol on an open square. The drones play as an "X" player, and the user is playing as an "O" player. The game is over if one of the players has three identical elements in a row: horizontally, vertically, or diagonally. The game can end with a draw result if there is no possibility to achieve any winning combination. The board is represented by two-dimensional matrix 3x3 (Fig. \ref{fig:board}), where each cell was enumerated as 1, 2, 3, ... 9. Each element of the matrix equals one of the following value:
0 : Unoccupied Cell; +1: Drone Symbol "X"; -1: Player symbol "O".

\subsection{Improved Basic Algorithm}

We can choose who makes the first turn in our game: the human or the drone swarm. The algorithm with a moderate level of difficulty is implemented, which allows the human player to win/lose equally.
Our developed algorithm for the Tic-tac-toe is based on the Basic Algorithm strategy introduced in \cite{Karamchandani}, which is focused on the computer-based game scenario.

Since the proposed drone-based scenario of Tic-tac-toe requires more preparation time and complex actions from the swarm, in this research we hypothesized that the less complexity of the game would not meet the player's expectations. To provide a considerable challenge for the user, we propose an Improved Simple Algorithm (see Algorithm 1).

\begin{algorithm} [h]
\caption{Improved Basic Algorithm}
\While {game is not over}
 {
    human turn\;
    
    checking for winning condition
    
    \eIf{drones can win (row, column, diagonal)}
    {
    insert "X" in the third cell and end game\;
    }
      {
        \eIf{human can win (row, column, diagonal)}
        {
        playing defensive\;
        }
        {
        play to make 2 in a row, column, diagonal (50 \%)\; 
        or random choice (50 \%)\; 
        }
      }
    {
    drone moving command\;
    
    checking for winning condition\;
    }
    \If {drones move first}
    {
    making the first move (50 \%)\;
    or random choice (50 \%)\;
    }
    \If {winning condition}
    {
    end game\;
    }
  }

\end{algorithm}

\section{Experiments}
  \subsection{Research Methodology}

\paragraph{Participants} 

We invited 13 participants aged 22 to 43 years (mean=25.6, std=5.9) to complete the survey. 15.4\% of them have never interacted with drones before, 15.4\% regularly deal with drones, and 69.2\%  periodically participate in drone-based scenarios.

\paragraph{Procedure} 
  
At the beginning of the experiment, the procedure and game equipment were introduced to each participant. Rules of Tic-tac-toe were described for 2 participants (15\%) who have never played the game before. The goal of Tic-tac-toe game is to make a line of 3 playing elements sooner than your opponent. Game elements, i.e., noughts for human-player and crosses for SwarmPlay, were represented by cardboard plates with printed black circles and cross shapes of the drones, respectively. Players placed the playing elements on the horizontally arranged game board, 1x1.2 m white board with grid lines. All participants had played two matches with SwarmPlay. A human made the first move in the first game and drones in the second. After each game, the game duration and score were recorded. 

At the end of the game, all respondents were asked to evaluate the SwarmPlay game with a Likert scale (1-5) on seven parameters: excitement, engagement, latency, challenge, tiredness, stress factor, and Turing test. 

  \subsection{Experimental results}

We conducted a chi-square analysis based on the frequency of answers in each category.

\begin{figure}
  \subfloat[ \label{fig_gen:1a}]{\includegraphics[width=1\linewidth]{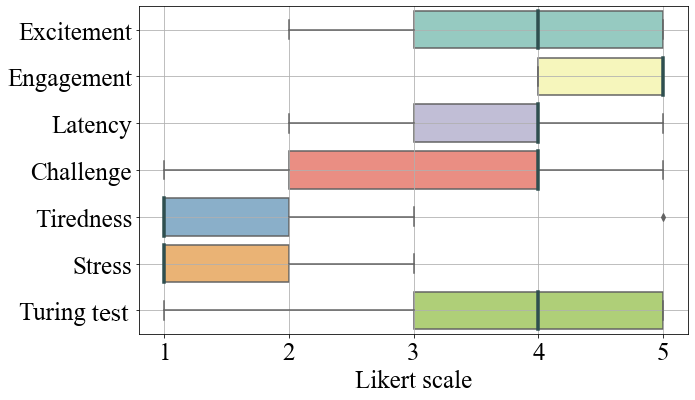}}\hfill
  \subfloat[ \label{fig_gen:1b}] {\includegraphics[width=0.9\linewidth]{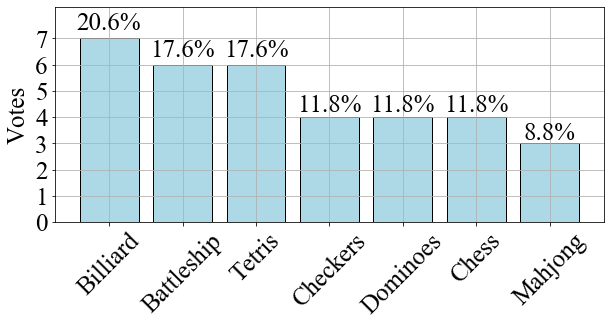}}\hfill
  \caption{Experimental results: a) Subjective feedback on 5-point Likert scale. b) Results of the survey: "What other games, based on SwarmPlay, would you like to play?"} \label{fig_gen}
\end{figure}

The results showed that the game parameters are all independent (min $p$ = 0.14 > 0.05). 
Additionally the chi-square test of independence revealed that the participants' experience with drones does not affect the evaluation of drone swarm perception criteria, such as tiredness ($\tilde{\chi}^2$=10.77, $p$=0.29), stress factor ($\tilde{\chi}^2$=2.53, $p$=0.87) and Turing test ($\tilde{\chi}^2$=12.19, $p$=0.20).
The results of the study are presented in
Fig. \ref{fig_gen:1a}.

In summary, 38.5\% of the participants found the game more exciting than regular paper-based game; 76.9\% did not feel any discomfort playing along with drones and 69.2\% of users found the SwarmPlay response fast enough ($\geq$ 4)
compared to the usual human-opponent move. The results revealed that participants were fully engaged in the game based on the Improved Simple Algorithm (69\% put 5 scores, 31\% - 4 scores), and 85\% claimed they did not get tired playing with drones ($\leq$ 2). The proposed algorithm proved itself comparable to average person skills, allowing its opponent to win or lose equally. Only 23\% of participants considered that playing with a robotic opponent was much distinguishable from the real person. 62\% of respondents evaluated the game being challenging ($\geq$ 4). 
Participants show a great interest to try other well-known board games in a new interpretation with a swarm of drones (Fig. \ref{fig_gen:1b}).

According to the results of the survey, the most popular games which people are willing to play with drones are Billiard with 20.6\% user choice, followed by Battleship (17.6\%) and Tetris (17.6\%). As the results from 26 games in total, SwarmPlay managed to win 23.1\% of all matches versus 26.9\% of the user wins. The average duration of one game was 67.2 sec. For the games that started with the SwarmPlay's move, the average time was 72.5 sec, which is 14.6\% longer than for the games with a human's first move (61.9 sec).

    \subsubsection*{Discussion} 

The results revealed that the first move is essential for the game outcome 
 (see Table \ref{tab:start-win}), with the first player winning in 46\% cases. According to the received data, SwarmPlay won 38.5\% of matches when drones started the game, while drones won only 7.7\% when a human player moved first. Draw outcomes occurred more frequently when SwarmPlay started (61.5\%), than when a human started the game (38.5\%). When human players started, they won in 53.8\% games.

\begin{table}[]
    \centering 
    \caption{Relation between first move and game results}
\begin{tabular}{|l|c|c|c|}
\hline
Result: $\textbackslash$ Start: & \begin{tabular}[c]{@{}c@{}}SwarmPlay \\ First move\end{tabular} & \begin{tabular}[c]{@{}c@{}}Human\\ First move\end{tabular} & \begin{tabular}[c]{@{}c@{}}Average time\\ (sec)\end{tabular} \\ \hline
SwarmPlay won & 5 & 1 & 65.1 \\ \hline
Draw & 8 & 5 & 69.7 \\ \hline
Human won & 0 & 7 & 58.6 \\ \hline
\textbf{Games in total:} & \textbf{13} & \textbf{13} & \textbf{67.2} \\ \hline
\end{tabular}
\label{tab:start-win}
\end{table}

Additionally, we discovered that the more sophisticated strategy SwarmPlay performed and the more points it had, the more human-like behaviour of the SwarmPlay participants mentioned.

\section{Conclusions and Future Work}

We have developed SwarmPlay, the system in which human plays against the swarm of drones. Our experimental results show that 38.5\% of the participants found the game more exciting than regular paper-based games, and 76.9\% of users did not feel stress while participating in the HDI scenario. 

Participants were engaged with the novel drone interaction technology (engagement mean score equals 4.7 out of 5.0) and indicated their readiness to play other drone-based games such as Billiard (20.6\% user choice), Battleship (17.6\%), and Tetris (17.6\%). 
Therefore, the proposed SwarmPlay technology can potentially improve our way of interaction with game pieces. Machines can learn from a human's winning strategy and, more importantly, teach humans how to achieve such a strategy throughout the interaction with an intelligent swarm.

The future work will be devoted to more advanced board games, and we plan to apply ML techniques to learn the level of the player and adjust the difficulty level of the game in real-time. 

\bibliographystyle{ACM-Reference-Format}


\end{document}